\documentclass[aps,prl,twocolumn,showpacs,superscriptaddress,groupedaddress]{revtex4}  
\usepackage{graphicx}  
\usepackage{dcolumn}   
\usepackage{bm}        
\usepackage{amssymb}   
\usepackage[latin1]{inputenc}
\usepackage{amsmath}
\usepackage{amsfonts}
\usepackage{amssymb}
\usepackage{setspace}
\usepackage{mathtools}
\usepackage{pstricks}
\usepackage{color}
\usepackage[english]{babel}
\usepackage{blindtext}
\usepackage{color}
\usepackage{hyperref}

\providecommand{\f}[2]{\frac{{#1}}{{#2}}}

\newcommand{\ee}[1]{\begin{equation}#1\end{equation}}
\newcommand{\ea}[1]{\begin{align}#1\end{align}}

\usepackage{accents}
\newcommand{\ubar}[1]{\underaccent{\bar}{#1}}
\begin{document}

\title{Curvature induced running of the cosmological constant}
\author{T.~Markkanen}\email{t.markkanen@imperial.ac.uk}\affiliation{Theoretical Physics, Blackett Laboratory, Imperial College, London, SW7 2AZ, United Kingdom}\affiliation{University of Helsinki and Helsinki Institute of Physics, Helsinki, P. O. Box 64, FI-00014, Finland}

\date{\today}

\begin{abstract}
In this work we investigate the renormalization group flow of the cosmological constant $\Lambda$ induced by the change in space-time curvature in the electroweak vacuum. We calculate the generic magnitude resulting from running in the standard model in a subtraction scheme that respects the Appelquist-Carazzone decoupling theorem. Interestingly, we find in this prescription that for a non-minimal coupling $\xi\lesssim  10^4$ the magnitude of the generated contribution remains below the value consistent with observations.
\end{abstract}
\maketitle
At present it appears that the Universe is undergoing accelerated expansion consistent with positive vacuum energy, or in other words a cosmological constant \cite{Ade:2013zuv}. Its extremely small value $\Lambda\sim 10^{-47} \,{\rm GeV}^4$ however, has yet to be satisfactorily explained as theoretical predictions suggest a value many orders of magnitude larger. This is the cosmological constant problem. In fact, there are two problems: The first issue concerns naturalness as the quantum zero-point energy implies a value around 120 orders of magnitude above observations \cite{Weinberg:1988cp}. It resembles the Higgs mass hierarchy problem and despite the existence of such a problem, for small energy-scales the standard model (SM) is well in accord with experiments \cite{Altarelli:2014xxa}. This leads one to question if at low energies the lack of naturalness is really a problem for $\Lambda$ and in this work we simply disregard this issue, despite its importance \cite{Nobbenhuis:2004wn}.

The second problem is different in nature and contrary to the first it is also present when regularizing dimensionally. This problem is due to electroweak (EW) symmetry breaking, which induces negative vacuum energy with a large magnitude even at the classical level and is made worse by quantum corrections \cite{Martin:2012bt}. The induced energy density is roughly 55 orders of magnitude larger than the observed value, which is sometimes quoted as the real magnitude of the cosmological constant problem \cite{Sola:2013gha}.

It has been argued that the induced vacuum energy may be set to zero as a result of a symmetry principle \cite{Witten:2000zk} or by treating it as a non-gravitating contribution \cite{Padmanabhan:2008if}. With such an assumption, the currently observed $\Lambda$ is often explained by some other mechanism, for which many viable models exist \cite{Copeland:2006wr}. However, this approach is made more complicated when quantum corrections for the SM are included: Any change in the energy scale induces a cosmological constant by renormalization group (RG) running \cite{ShapSol0,Ward:2009wq}.
Superficially, a massive particle of type $i$ gives a contribution to the running $\propto m_i^4$, which results in a value much larger than what is consistent with observations. A generically large RG induced contribution would imply that $\Lambda$ is very sensitive to the choice of renormalization condition: Fixing $\Lambda$ at some scale to be of observable magnitude would at current scales result in a value much different from observations, unless the fixing occurs very close to the scale where $\Lambda$ is measured. It has been argued that heavy particles should decouple from the running due to the Appelquist-Carazzone theorem \cite{Appelquist:1974tg} but finding the suitable renormalization prescription has proven to be challenging \cite{ShapSol,ShapSol2,Babic:2001vv}.

In this work we focus on the contribution to $\Lambda$ from RG running induced by the change in space-time curvature, $R$, from a top-down quantum field theory perspective. Related work can be found in \cite{works}. 
In our calculation thermal effects are neglected, which indicates that our results cannot be extended beyond the EW vacuum since electroweak symmetry breaking is fundamentally a finite temperature effect. Hence in terms of $R$, we must have $R\leq R_{\rm EW}$. 
In this region we can treat space-time as a classical background and assume Einstein gravity to be valid, if only as an effective theory. We will assume a state with vanishing expectation values for matter field fluctuations making $R$ the dynamical variable. With these assumptions we can use the framework of quantum field theory on a curved background \cite{ParkerTomsBirrelDavies} while acknowledging that a full quantum gravity treatment can change the picture \cite{Mottola}. 
Our conventions will be (+,+,+) in the classification of \cite{Misner:1974qy}, with $\hbar\equiv1\equiv c$.

We start by deriving the running $\Lambda$ in the modified minimal subtraction procedure $\overline{\rm MS}$, for a model with one non-minimally coupled scalar field and one Dirac fermion in the non-interacting case. Our action $S\equiv\int\sqrt{-g}\mathcal{L}$ will then have the familiar Einstein-Hilbert part \footnote{It is well-known that quantum corrections in curved space generate terms of order $R^2$. We treat such effects negligible, not to be confused with constants $\sim R_{\rm EW}^2$.}
\ee{\mathcal{L}
=\mathcal{L}_g+\mathcal{L}_m=-\Lambda+(2\kappa)^{-1}{R}+\mathcal{L}_m\, ,\label{eq:act0}}
and a matter part
\ea{\mathcal{L}_m
=&-\f{1}{2}(\nabla_\mu \phi)^2-\f{1}{2}m_\phi^2\phi^2-\f{1}{2}\xi R\phi^2\nonumber \\&-\bar{\psi}(i\underline {\gamma}^\mu\nabla_\mu+m_\psi)\psi\, ,\label{eq:act}}
with the standard curved space generalizations of the Klein-Gordon and Dirac Lagrangians \cite{ParkerTomsBirrelDavies} and where $\kappa^{-1}\equiv(8\pi G)^{-1}\equiv M_{\rm pl}^2$. Throughout this work we will assume that gravitational dynamics are given by some unspecified classical fluid that dominates over the quantum backreaction. This means that effectively we can treat $R$ as a free parameter. Deriving the effective potential to one-loop order can be found in \cite{CW} and its curved space generalization for small curvature in \cite{Parker:1984dj}, giving 
\ea{&\mathcal{L}_{\rm eff}=-\Lambda+\f{R}{2\kappa}-\f{1}{2}m_\phi^2\phi^2-\f{1}{2}\xi R\phi^2\nonumber \\&-\f{M_\phi^4(R)}{64\pi^2}\bigg[\log\bigg(\f{M_\phi^2(R)}{\mu^2}\bigg)-\f{3}{2}-\Big\{\f{2}{\epsilon}-\gamma_e+\log(4\pi)\Big\}\bigg]\nonumber \\&+\f{M^4_\psi(R)}{16\pi^2}\bigg[\log\bigg(\f{M_\psi^2(R)}{\mu^2}\bigg)-\f{3}{2}-\Big\{\f{2}{\epsilon}-\gamma_e+\log(4\pi)\Big\}\bigg]\, ,\label{eq:LQ}}
for a constant $\phi$. The scale parameter introduced by dimensional regularization is $\mu$, $n=4-\epsilon$ is the number of dimensions and the effective masses are
\ee{M_\phi^2(R)=m_\phi^2+R\bigg(\xi-\f{1}{6}\bigg)\, , \quad M_\psi^2(R)=m_\psi^2+\f{R}{12}\, .\label{eq:effM}}
It should be noted that in (\ref{eq:LQ}) the notation $\phi$ really means the expectation value of the quantized field, $\langle\hat{\phi}\rangle$. In the usual $\overline{\rm MS}$ renormalization scheme one chooses the counter terms to cancel all the contributions within the curly brackets in (\ref{eq:LQ}). In this letter we will also include the term $-3/2$ in the $\overline{\rm MS}$ subtraction
. Assuming a state with $\phi=0$, $R$ will be the only dynamical variable of the problem. Renormalization group improvement for the curved space expression (\ref{eq:LQ}) comes via the Callan-Symanzik equation just like in flat space:
\ee{\bigg\{\mu\f{\partial}{\partial\mu}+\beta_{c_i}\f{\partial}{\partial{c_i}}-\gamma \phi\bigg\}\mathcal{L}_{\rm eff}=0\,, \qquad\beta_{c_i}\equiv \mu\f{\partial c_i}{\partial\mu}\, ,\label{eq:CS}}
where the $c_i$ stands for all the parameters of the action with summation over the repeated index $i$ assumed. With no interactions, the only non-zero $\beta$-functions are the ones for $\kappa^{-1}$ and $\Lambda$. Deriving the $\beta$-functions in $\overline{\rm MS}$ is a standard calculation \cite{Toms:1983qr} and gives the results
\ee{\bar{\beta}_{\kappa^{-1}}=-\f{m^2_\phi(\xi-{1}/{6})-m^2_\psi/{3}}{8\pi^2}\, , \quad \bar{\beta}_\Lambda=\f{m_\phi^4-4 m^4_\psi}{32\pi^2}\,,\label{eq:msb}}
where the overline is used to denote an $\overline{\rm MS}$ quantity. For $\beta$-functions to be calculated in a different scheme we will use an underline, $\ubar{\beta}$. 

Essentially, the Callan-Symanzik equation (\ref{eq:CS}) is an expression of invariance with respect to the renormalization scale, $(d/d\mu)\mathcal{L}_{\rm eff}=0$, so after substituting the solutions of (\ref{eq:CS}) back into (\ref{eq:LQ}) we can in principle pick $\mu$ freely. However our perturbative result is only correct up to higher order corrections and not completely $\mu$-invariant, so one should choose $\mu$ such that the neglected corrections are as small as possible, as discussed in \cite{Ford:1992mv}. In practice this means that for example in the result (\ref{eq:LQ}), we should choose $\mu$ so that the logarithms remain small for all $R$, i.e in curved space $\mu$ should become a function of $R$ \cite{Herranen:2014cua}. This step is crucial, since it will introduce a natural suppression to the result. The choice that makes the logarithms vanish we will call the optimal choice and denote it as $\mu(R)$. 
For example, if we have only a scalar, 
$\mu^2(R) = M^2_\phi(R)$ and similarly $\mu^2(R) = M^2_\psi(R)$ if we only have a fermion. With both a scalar and a fermion we can choose
\ea{\log\mu^2(R)&=\f{M^4_\phi(R)\log M^2_\phi(R)-4M^4_\psi(R)\log M^2_\psi(R)}{M^4_\phi(R)-4M^4_\psi(R)}\nonumber \\&=\f{\sum_i n_i M^4_i(R)\log M^2_i(R)}{\sum_in_i M^4_i(R)}\equiv Y(R)\,,\label{eq:opscale}}
where $M_i$ is the effective mass of a particle type $i$ and $n_i$ counts the degrees of freedom. The optimal scale also respects the general conditions advocated in \cite{Babic:2004ev} \footnote{From $\kappa^{-1}(R),\Lambda(R)$ we can easily derive a generalized Einstein equation with running parameters, which was the basis for the derivations in \cite{Babic:2004ev}.}.

Using (\ref{eq:opscale}), the RG improved effective action has precisely the same form as the Einstein-Hilbert action (\ref{eq:act0}), but with the quantum corrections manifesting themselves as running of the parameters: $\kappa^{-1}, \Lambda\Rightarrow\kappa^{-1}(R),\Lambda(R)$. 
In this work we are only interested in the generic magnitude for the contribution from RG running to $\Lambda$, which essentially means the difference when evaluated at two scales. In such an approach one can neglect the difficult question of how to define the physical value. If we set $\Lambda(R_{\rm EW})=0$ then the value at $\Lambda(0)$ will correspond to the maximal contribution generated by running. 
From (\ref{eq:msb}) we can solve  $\Lambda(R)$ in the $\overline{\rm MS}$ scheme
\ea{\Lambda(R)&=(R-R_{\rm EW})\f{m_\phi^4-4 m^4_\psi}{64\pi^2}Y'(0)+\mathcal{O}(R_{\rm EW}^2)\nonumber\\\Leftrightarrow~~\Lambda(R) &=\begin{dcases}&(R-R_{\rm EW})\f{m_\phi^2(\xi-1/6)}{64\pi^2},\quad  m_\psi= 0\\&(R_{\rm EW}-R)\f{m_\psi^2}{192\pi^2},\quad\quad\quad\quad  m_\phi= 0\end{dcases}\label{eq:MSLambda}}
In (\ref{eq:MSLambda}) we can see that contrary to the naive assumption, a term $\sim m_i^4$ in the $\bar{\beta}_\Lambda$-function (\ref{eq:msb}) induces a contribution $\sim m_i^2R_{\rm EW}$ for the cosmological constant. This is because minimizing the higher loop corrections gives $\mu$ an $R$ dependence via $\mu(R)$. However, the result is still very large. From \cite{Caldwell:2013mox} we have a rough estimate $R_{\rm EW}\sim \kappa T_{\rm EW}^4$ with $T_{\rm EW}$ being the EW transition temperature. Assuming a single fermion with  $m_\psi\sim T_{\rm EW}\sim 2\times10^2\,{\rm GeV}$ we have $\Lambda(0)\sim 10^{-26}\,{\rm GeV}^4$. This 
implies that RG running generically results in a contribution with an absolute value 21 orders of magnitude larger than what is consistent with observations. Fortunately, this is not the whole story, as there are severe issues associated with our use of $\overline{\rm MS}$ in a low energy regime.

The problems with $\overline{\rm MS}$ can be made apparent with a simple example. Supposing flat space and a Yukawa interaction term in (\ref{eq:act}), $\mathcal{L}_I = -g \phi \bar{\psi}\psi$,
we can parametrize the full interacting propagator for the scalar field in momentum space with the effective mass parameter $\Sigma^2_p$ as
\ea{iG(p)&=\int d^4x~i\langle\hat{\phi}(x)\hat{\phi}(0)\rangle e^{-ip\cdot x}\nonumber =\big(p^2+m_\phi^2+\Sigma^2_p\big)^{-1}\, ,}
in the normalization conventions of \cite{Peskin:1995ev} for the field operators and Fourier space. The relevant counter terms included in $\Sigma^2_p$ are $p^2\delta Z$ and $\delta m_\phi^2$ and in our version of $\overline{\rm MS}$ we choose them to subtract the divergent pole and all scale independent numbers such that the  one-loop result for $\Sigma^2_p$ is
\ea{-i\Sigma^2_p&=\f{3ig^2}{4\pi^2}\int^1_0dx~\big[m_\psi^2+x(1-x)p^2)\big]\nonumber \\&\times\log\bigg(\f{m_\psi^2+x(1-x)p^2}{\mu^2}\bigg)\, .\label{eq:M}}
where $\mu$ is again introduced by dimensional regularization. $G(p)$ also satisfies a Callan-Symanzik equation
\ea{\bigg\{\mu\f{\partial}{\partial\mu}+\bar{\beta}_{m^2_\phi}\f{\partial}{\partial m^2_\phi}+2\gamma \bigg\}G
(p)&=0\nonumber \\\Leftrightarrow\quad\mu\f{\partial}{\partial\mu}\Sigma^2_p
+\bar{\beta}_{m^2_\phi}-2\gamma(p^2+m^2_\phi) &=0+\mathcal{O}(\hbar^2)\, ,\label{eq:CSp}}
where the higher order corrections are denoted as $\mathcal{O}(\hbar^2)$. The solution gives the $\overline{\rm MS}$ $\beta$-function for $m^2_\phi$
\ee{\f{\bar{\beta}_{m^2_\phi}}{m_\psi^2}=\frac{3 g^2}{2 \pi ^2}\bigg(\f{m^2_\phi}{6m^2_\psi}-1\bigg)\, .}
However, doing the above calculation in a different subtraction scheme will reveal important features that are also crucial for the running of $\Lambda$. One could equally well do the calculation in a more physical renormalization scheme where one includes in $p^2\delta Z$ and $\delta m_\phi^2$, in addition to scale independent numbers, also the logarithm in (\ref{eq:M}) evaluated at $p^2=-\tilde{\mu}^2$. This will give the effect of replacing $\mu^2$ in (\ref{eq:M}) with $\mu^2(\tilde{\mu})=m_\psi^2-x(1-x)\tilde{\mu}^2$. Solving (\ref{eq:CSp}), now with $\tilde{\mu}$ being the renormalization scale, will give the $\beta$-function in this scheme
\ea{\f{\ubar{\beta}_{m^2_\phi}}{m^2_\psi}&=\f{3g^2}{2\pi^2}\int^1_0dx~\frac{\tilde{\mu}^2 x(1-x) \left[m_\psi^2 -x(1-x)m_\phi^2\right]}{m^2_\psi\left(m_\psi^2-x(1-x) \tilde{\mu}^2 \right)}\nonumber \\&=\begin{dcases}&\frac{3 g^2}{2 \pi ^2}\bigg(\f{m^2_\phi}{6m^2_\psi}-1\bigg)+\mathcal{O}\bigg(\f{m_\psi^2}{\tilde{\mu}^2}\bigg),\quad ~~~\,m_\psi< \tilde{\mu}\\&\frac{g^2 \tilde{\mu}^2}{4 \pi ^2 m_\psi^2}\bigg(1-\f{m^2_\phi}{5m^2_\psi}\bigg)+\mathcal{O}\bigg(\f{\tilde{\mu}^4}{m_\psi^4}\bigg),~~  m_\psi> \tilde{\mu}\, .\end{dcases}\label{eq:YB}}
From (\ref{eq:YB}) we can see that $\ubar{\beta}_{m^2_\phi}$ coincides with the $\overline{\rm MS}$ result only at the limit when the mass of the fermion is much smaller than $\tilde{\mu}$. Conversely, when we are in the region where $\tilde{\mu}$ is much smaller than the mass,
$\ubar{\beta}_{m^2_\phi}$ has very different behaviour to the $\overline{\rm MS}$ result, effectively, the particle decouples. This is a manifestation of the Appelquist-Carazzone decoupling theorem \cite{Appelquist:1974tg}.

If we assume a similar effect also for the physical $\beta_\Lambda$-function renormalized in position space, for small renormalization scale it follows that we cannot expect the $\overline{\rm MS}$ results (\ref{eq:msb}) to be reliable \cite{ShapSol0,ShapSol,ShapSol2,Babic:2001vv}. Here, $R$ is the dynamical variable, much like $p^2$ was in the Yukawa-theory example and the renormalization scale will be $R$ evaluated at some point, $R_0$. For $\overline{\rm MS}$ to be valid we should have $R_0\geq m_i^2$, which is not true for many degrees of freedom of the SM.
Hence, we need to find a more suitable subtraction scheme.

First we state what we are after. Our assumption is that the running of $\Lambda$ is not correctly described by $\overline{\rm MS}$ at $R_0\rightarrow 0$, since in $\bar{\beta}_\Lambda$ heavy particles do not decouple. Furthermore, we want our $\beta$-function to coincide with the $\overline{\rm MS}$ result in the limit of large $R_0$. As emphasized in \cite{ShapSol2}, these conditions are met if we have
\ee{ \ubar{\beta}_\Lambda =\begin{dcases} \bar{\beta}_\Lambda,\quad R_0\rightarrow\infty \\
0,\quad~~ R_0\rightarrow0
\end{dcases}\, .
} 

To start, replace $\mu$ with $\mu(R_0)$, which is 
at this stage only a parametrization. 
Note that if we want to minimize the higher order corrections, this parametrization must be done via the optimal scale (\ref{eq:opscale}). 
Next we include a counter term  as $\Lambda\rightarrow\Lambda+\delta\Lambda\big(\mu(R_0)\big)$. 
The equation for $\ubar{\beta}_\Lambda$ from (\ref{eq:msb}) is then
\ee{\mu(R_0)\f{\partial\big[\Lambda+\delta\Lambda\big(\mu(R_0)\big)\big]}{\partial \mu(R_0)}=\f{\sum_i n_i M^4_i(0)}{32\pi^2}
\, .\label{eq:betatemp}}
Assuming that we can express the $\ubar{\beta}_\Lambda$-function as a Laurent series in $\mu(R_0)$, we can write the contribution from the counter term as
\ee{\mu(R_0)\f{\partial\,\delta\Lambda\big(\mu(R_0)\big)}{\partial \mu(R_0)}=\f{\sum_i n_i M^4_i(0)}{32\pi^2}\sum_{k>0} a_k\bigg(\f{\mu^2(0)}{\mu^2(R_0)}\bigg)^k
\, .\label{eq:subf}}
In the above we have chosen all coefficients with $k\leq0$ to vanish as we want a result that coincides with $\overline{\rm MS}$ for large $\mu(R_0)$ and further imposed that the result contains only integer powers of $\mu^2(R_0)$. Requiring a vanishing $\ubar{\beta}_\Lambda$-function at $R_0\rightarrow 0$ gives the constraint $\sum_{k>0}a_k =1$.
The precise values of the $a_k$ can be determined by choosing a particular renormalization scheme, preferrably motivated by a physical process. Our prescription will be to simply include only the leading term by setting $a_1=1$ to be the only non-zero coefficient, giving \footnote{Solving for $\delta\Lambda\big(\mu(R_0)\big)$ from (\ref{eq:subf}) shows that the corresponding counter term can always be chosen such that it never introduces $\mathcal{O}(m_i^4)$ terms at the level of the action.}
\ee{\mu(R_0)\f{\partial\Lambda}{\partial \mu(R_0)}=\f{\sum_i n_i M^4_i(0)}{32\pi^2}\bigg(1-\f{\mu^2(0)}{\mu^2(R_0)}\bigg)\equiv\ubar{\beta}_\Lambda\, .\label{eq:betatemp2}}
From (\ref{eq:betatemp2}) one may see the correct behaviour of the subtraction term: Negligible at large $R_0$ while giving decoupling at $R_0=0$. Again setting $\Lambda(R_{\rm EW})=0$, we have
\ee{\Lambda(R)=\f{\sum_i n_i M^4_i(0)(R^2-R_{\rm EW}^2)}{128\pi^2}\big[Y'(0)\big]^2+\mathcal{O}(R_{\rm EW}^3)\, .\label{eq:physL}}
Since $Y'(0)\propto m^{-2}$ in (\ref{eq:physL}) we see that running generically results in a small contribution. Even if one does not choose a particular prescription, the decoupling requirement ensures that the first potentially non-zero term in (\ref{eq:physL}) is $\mathcal{O}(R^2_{\rm EW})$ \footnote{A direct computation using (\ref{eq:subf}) shows that the leading term will be the right hand side of (\ref{eq:physL}) multiplied by $\sum_{k>0}a_k k$.}. Our method of finding a counter term via a Laurent series in (\ref{eq:subf}) is very similar to the approach of \cite{Babic:2001vv}, the main difference being that here $\mu$ becomes a function of the renormalization point $R_0$ due to the minimization of the quantum corrections.

As an example, for the action in (\ref{eq:act}) with $|\xi|\lesssim10^6$ and for simplicity setting  $m_\phi=m_\psi$ we get the result
\ea{\Lambda(R)&=\big(R_{\rm EW}^2-R^2\big)\f{\big[1/3-(\xi-1/6)\big]^2}{384\pi^2}+\mathcal{O}(R_{\rm EW}^3)
\nonumber \\&\Rightarrow\quad \Lambda(0)\lesssim 10^{-47}{\rm GeV}^{4} \, ,\label{eq:lambex}}
where we again 
used $R_{\rm EW}\sim \kappa T_{\rm EW}^4$ with $T_{\rm EW}\sim 2\times 10^2\,{\rm GeV}$ from \cite{Caldwell:2013mox}. 
An intriguing result showing that running naturally generates contributions with magnitudes smaller or comparable to what is consistent with current observations. A large value $\xi\sim 10^6$ can seem extreme, but in fact for the Higgs field the current observational bound is $\xi\leq 2.6\times 10^{15}$ \cite{Atkins:2012yn}. 

Now we derive a result for the relevant degrees of freedom of the SM. 
As a first approximation if we assume no running for the SM mass parameters, we can directly use the formula (\ref{eq:physL}). The factor $\sum_i n_i M^4_i(0)$ simply sums all degrees of freedom and can be trivially obtained from the SM Lagrangian. To a good approximation we can include only the most massive particles, the $W^\pm$ and $Z^0$ bosons, the top quark and the Higgs particle \footnote{The effective masses for the Goldstone bosons in the 't Hooft-Landau gauge gives a term  proportional to $ R^3_{\rm EW}/m^2\log (R_{\rm EW}/m^2)$ in the end result, which we treat as negligible.}. The complicated step is obtaining the optimal scale for the SM, $Y_{\rm SM}(R)$ in (\ref{eq:opscale}), since the effective masses come with different $R$-dependences \cite{Herranen:2014cua}. However, if (\ref{eq:lambex}) is any indication, we may expect that only for large $\xi$ we have a $\Lambda(0)$ comparable with observations. Hence we only calculate the leading term in an expansion in powers of $\xi^{-1}$. 

For $h$ that is the real part of the Higgs doublet that acquires an expectation value, we have the Lagrangian
\ee{\mathcal{L}_h=-\f{1}{2}(\nabla_\mu h)^2+\f{1}{2}m^2h^2-\f{1}{2}\xi R h^2-\f{\lambda}{4}h^4\, ,}
and because $m^2>0$ it has a non-zero vacuum expectation value $\langle\hat{h}\rangle^2 ={({m^2-\xi R})/{\lambda}}$.
Since all particles get their masses via interaction with $h$, they will also have coupling to $R$ via $\xi$, which allows us to parametrize all effective masses in terms of $m$ and $\xi R$. This greatly simplifies the expression for $Y_{\rm SM}(R)$ in (\ref{eq:opscale}):
\ea{M^2_i(R)&= a_i\langle\hat{h}\rangle^2 +b_iR
\approx a_i(m^2-\xi R)\nonumber \\ \Rightarrow\quad Y_{\rm SM}'(0)&=-\f{\xi}{m^2}+\mathcal{O}(1/\xi)^0 \, ,\label{eq:par}} 
where $b_i$ is a factor intrinsic to a particle type $i$ and $a_i$ comes from the Higgs interaction \footnote{To be precise the $M_i$ for gauge fields have dependence on $R_{\mu\nu}$ \cite{Herranen:2014cua}. However, for the leading term in a small curvature expansion these terms sum to $R$, a consequence of general covariance. 
}. Using (\ref{eq:par}) 
and writing all masses in terms of $m$ we get from (\ref{eq:physL})
\ea{\Lambda_{\rm SM}(R)&=\f{R_{\rm EW}^2-R^2}{128\pi^2}\xi^2\nonumber \\&\times\frac{48 y^4-9 g_1^4-6 g_1^2 {g_2}^2-3 {g_2}^4-64 \lambda^2}{16 \lambda^2}
\,, \label{eq:lambdasm}}
in the conventions of  \cite{Herranen:2014cua} where $g_1$, $g_2$, $y$ are the SU(2), U(1) and top Yukawa couplings and we have neglected $\mathcal{O}(\xi,R_{\rm EW}^3)$ corrections. 
Using the values of the $\overline{\rm MS}$ couplings at the EW scale from \cite{Buttazzo:2013uya}, we see that only for  $\xi\sim 10^4$ we start to approach a contribution of the observed magnitude. Since the above contribution corresponds to the maximal value induced by running, we can conclude that as long as $\xi\lesssim 10^4$ also in the SM running induces only a small correction to the cosmological constant.
As the top quark dominates (\ref{eq:lambdasm}), essentially the right result is obtained by retracing the steps with just a single fermion.

What we have shown is that renormalization group flow in the electroweak vacuum induced by changing space-time curvature gives only small corrections to the cosmological constant, when the non-minimal coupling is smaller than $\mathcal{O}(10^4)$. Our approach is based on the well-established approaches of renormalization group improvement and curved space field theory. 
One reason for this result is that in curved space minimizing the higher order corrections requires $\mu$ to be a particular function of $R$ resulting in a natural suppression. Additionally, we defined a new subtraction scheme for the $\beta$-function of $\Lambda$. We motivated this scheme by assuming that like in other areas of particle physics the $\overline{\rm MS}$ scheme fails at small scales due to the decoupling theorem. We note that a more desirable route for justifying this subtraction would be via renormalization of a physical process. An intriguing detail is that $\xi\sim 10^4$ gives a contribution comparable to the observational value of $\Lambda$. However, a thorough investigation including interactions and temperature in a physically motivated subtraction scheme is needed before conclusions can be reached. Temperature corrections are also important for the intricacies during EW symmetry breaking and it is likely that for $R\geq R_{\rm EW}$ non-trivial effects also for the running of $\Lambda$ arise. An investigation along the lines of \cite{newstudy} is also needed to determine if the running of $\Lambda$ is in accord with other aspects of cosmology such as the formation of structure. Similarly, a running $\Lambda$ may play an important role in early universe physics such as inflation and its graceful exit, as discussed in \cite{Sola:2015rra}.

\acknowledgments{TM would like to thank Robert Brandenberger, Matti Herranen, Mark Hindmarsh, Sami Nurmi, Arttu Rajantie, Kari Rummukainen, Joan~Sol\`{a}, Anders Tranberg and Aleksi Vuorinen for discussions, the University of Barcelona, the University and Sussex and McGill University for hospitality and the Mikael Bj\"{o}rnberg Memorial Fund and the Oskar Huttunen Foundation for support.}

\end{document}